\documentclass[submission,copyright,creativecommons]{eptcs}
\providecommand{\event}{VPT 2020} %
\usepackage{breakurl}             %
\usepackage{underscore}        

\usepackage{graphicx}
\usepackage{galois}
\usepackage{stackrel}

\usepackage{amsmath} 
\usepackage{algorithm}
\usepackage[noend]{algpseudocode}

\usepackage{alltt}
\usepackage{tikz}
\usepackage{etex}
\usepackage[etex=true,export]{adjustbox}
\usepackage{framed} %

\usepackage{amssymb}
\usepackage{upquote}
\usepackage{fancyvrb} %

\usepackage{float}
\usepackage{wrapfig}

\usepackage{hyperref}
\usepackage{listings} 
	   \usepackage{courier}
\usepackage{paralist}
\usepackage{pdfpages}

\newcommand{\anydim}[1]{\ge\! 0}

\def\ll{[\![}
\def\rr{]\!]}

     {\end{tabular}\end{tt}\end{small}}
\def\anno#1{{\ooalign{\hfil\raise.07ex\hbox{\small{\rm #1}}\hfil%
        \crcr\mathhexbox20D}}}

\title{An Experiment Combining Specialization \\with Abstract Interpretation}
\author{John P. Gallagher \institute{Roskilde University, Denmark \and IMDEA Software Institute, Spain}  \email{jpg@ruc.dk} 
\and
Robert Gl\"{u}ck \institute{Copenhagen University, Denmark} \email{glueck@acm.org}
}

\def\titlerunning{An experiment in specialising an abstract interpreter}
\def\authorrunning{J. P. Gallagher and R. Gl\"{u}ck}

\begin{document}
\maketitle

\pagestyle{plain}
\pagestyle{myheadings}

\paragraph{Introduction and motivation.}
It was previously shown that control-flow refinement can be achieved by a program specializer incorporating property-based abstraction, as described in 
\cite{Gallagher-VPT2019} and applied in \cite{DBLP:journals/tplp/DomenechGG19} to improve termination and complexity analysis tools.
We now show that this purpose-built specializer can be reconstructed in a more modular way, and that the 
previous results can be 
achieved using an off-the-shelf partial evaluation tool, applied to an abstract interpreter. The key feature of the
abstract interpreter is the abstract domain, which is the product of the property-based abstract domain with the concrete domain.  This language-independent framework provides a practical approach to 
implementing a variety of powerful specializers, and contributes to a stream of research on using interpreters and specialization to achieve program transformations.

\textbf{Abstract interpreters.}
Let $L$ be a programming language.
We consider a program $p \in L$ to be a partial function of one argument (possibly an $n$-tuple), denoted $\ll p \rr $, and assume
that both argument and result are elements of a set $S$.  %
An \emph{interpreter} for $L$ is a program $I$ such that
for all $p \in L$, $v \in S$, $\ll I \rr(p~ v) = \ll p\rr~v$ (or both are undefined).  
An \emph{abstract interpreter} computes
a \emph{safe approximation} of an interpreter $I$.
For the present discussion, we say that
an abstract interpreter $A$ takes a program $p \in L$ with a \emph{set} $\phi$ of input values, and computes a
set of values as output. $A$ is a safe approximation of $I$
if for all $p \in L$ and $\phi\in \wp(S)$, $ \{\ll I \rr (p~v) \mid v \in \phi\} \subseteq \ll A\rr(p ~\phi)$.
In other words, $A$ over-approximates the set of results that $I$ computes on elements of $\phi$.

In practice, abstract interpreters 
represent elements of $\wp(S)$ by descriptions in some \emph{abstract domain} $D$.
 If $D$ is finite,  an abstract interpreter using $D$ is a total function $\ll A \rr: (L \times D) \rightarrow D$, that is, $\ll A \rr(p~\phi)$ terminates for all $p \in L$ and $\phi \in D$. 
Although abstract interpreters are typically designed to
terminate,  domains that are infinite and abstract
interpretations that do not
guarantee termination are also useful; we can still
gain interesting information from running them.
In what follows, we define a \emph{mixed} abstract interpreter that
combines the concrete domain $\wp(S)$ with a finite domain $D$.
In the theory of abstract interpretation \cite{Cousot-Cousot},
the Cartesian product $\wp(S) \times D$ is an abstract
domain.
The abstract interpreter $A$ used in our experiment has type 
$(L \times ((\wp(S) \times D)) \rightarrow (\wp(S) \times D)$.

\textbf{Structure of an interpreter.}
Let us assume that an (abstract) interpreter operates as a transition system, though this is not essential.  A state consists of a point in the program being interpreted together with the values of variables in the domain of interpretation at that point. 
For the standard interpreter, let $q$ and $q'$ be program points, and $v,v' \in S$ be the respective values of the variables at those points. A transition is 
$\langle q, v \rangle \xrightarrow{\delta(v,q)=v'} \langle q', v' \rangle$,
where $\delta$ is a function relating $v$ and $v'$ at the point $q$.  A transition in an abstract interpreter over domain $D$ uses a mapping $\delta_D(\phi,q) = \phi'$ where $\phi,\phi' \in D$ ($\delta_D$ is sometimes called the abstract ``transfer function").

  Transitions in an abstract interpreter over the product domain $\wp(S) \times D$ have two components corresponding to $\wp(S)$ and $D$ respectively: they have the form
 $\langle q, \{v\}, \phi \rangle \xrightarrow{\delta(v,q)=v'; \delta_D(\phi,q)=\phi} \langle q', \{v'\}, \phi' \rangle$, where $\delta$ and $\delta_D$ are the transfer functions for $\wp(S)$ and $D$ respectively. Assuming that $\delta$ is the standard interpreter transfer function, this both computes
 the standard result as well as an abstract result in $D$.  We will see that we can exploit the separate components during specialization.
We assume that the initial call to the interpreter contains a singleton set $\{v\} \in \wp(S)$, and thus only singleton concrete states are reachable.

 \textbf{Specialization.}
 A \emph{specializer} for $L$ is a program $S$ that transforms a program $p \in L$ with respect to partially specified input.
We assume that $p$'s argument is a pair $(v_1~v_2)$ and that $S$ is provided with $v_1$.  The result is a program $p' \in L$, i.e. 
$\ll S \rr (p~v_1) = p'$ which satisfies the property  $\ll p' \rr ~v_2 = \ll p \rr (v_1~v_2)$. 
Specialization of an interpreter with respect to a program in $L$ is known as the \emph{first Futamura projection}~\cite{Futamura,Jones-Gomard-Sestoft}. We have
$\ll S \rr (I~p) = I_p$, where according to the properties of interpreters and specializers, $\ll I_p \rr v = \ll I \rr (p~v) =   
 \ll p\rr v$.  The program $I_p$ can be seen as the compilation or transformation of $p$ into the language of $I$.
  Values encountered during specialization are \emph{static} or \emph{dynamic},  in the terminology of partial evaluation. Functions with static arguments can be evaluated during specialization, while functions with dynamic arguments are not, and are retained in the specialised program. A \emph{binding-time analysis} \cite{Jones-Gomard-Sestoft} can determine which parts of the program to be specialised are guaranteed to be static.  

\textbf{Abstract interpreter specialization.}
Consider the specialization of an abstract interpreter with domain $\wp(S) \times D$. In a state $\langle q, \{v\}, \phi \rangle$, we can determine that the program point $q$ and the  abstract state $\phi$ are static, while $v$ is dynamic.  This is because the initial abstract state is static (even if it is the ``top" element of $D$) and in a transition from $\langle q, \{v\}, \phi \rangle$, 
where $q$ and $\phi$ are static, and $v$ dynamic,  $\delta_D(\phi,q)=\phi'$ can be evaluated while $\delta(v,q)=v'$ cannot; thus the computation $\delta(v,q)=v'$ is retained in the residual specialised interpreter.  Thus in the next state $\langle q', \{v'\}, \phi' \rangle$, $q'$ and $\phi'$ are static, while $v'$ is dynamic. Furthermore, if $D$ is finite, then the static values have \emph{bounded static variation}, which means that only a finite number of different values of the static arguments arise during specialization.
 This leads to a so-called \emph{polyvariant} specialization.
A transition of the form $\langle q, \{v\}, \phi \rangle \xrightarrow{\delta(v,q)=v'; \delta_D(\phi,q)=\phi'} \langle q', \{v'\}, \phi' \rangle$ is specialised into a finite number of transitions of the form
 $\langle q_{\phi}, v \rangle \xrightarrow{\delta(v,q)=v' } \langle q'_{\phi'}, v' \rangle$, for each pair $(q,\phi)$ encountered during specialization. 
 
 \textbf{Control-flow refinement by mixed interpreter specialization.}
 A abstract interpreter for constrained Horn clauses was written as a Prolog program\footnote{Available at \url{https://github.com/jpgallagher/absint4pe}}. 
 The abstract domain is a product domain as described above,
 where $D$ is a set $\wp(\Psi)$ where $\Psi$ is
 finite set of properties.
 The main interpreter predicate is \texttt{solve(Q,A,Phi,Psi,Prog)}, 
 representing a state of the interpreter with a call to predicate \texttt{Q} with concrete arguments \texttt{A}, abstract state (the set of properties \texttt{Phi} from \texttt{Psi} that \texttt{A} entails),  \texttt{Psi} and \texttt{Prog}, the last two being the set of all properties and the set of Horn clauses respectively.
 A transition of the interpreter evaluates two calls \texttt{delta} and \texttt{delta_D}, corresponding to
 $\delta$ and $\delta_D$ above. $\delta$ evaluates the constraints of the clauses, and the
 $\delta_D$ computes the properties for the body calls in the clause. When run normally, the interpreter mirrors the standard semantics, but in addition carries around the set of properties that hold.
 Consider the following example clauses considered in 
 \cite{Gallagher-VPT2019}.
 \begin{tabular}{l|l}
\begin{lstlisting}
while0(X,Y,M)$\leftarrow$ X>0,if0(X,Y,M).
while0(X,Y,M)$\leftarrow$ X=<0.
\end{lstlisting}
~&~
\begin{lstlisting}
if0(X,Y,M)$\leftarrow$ Y<M,Y1=Y+1, while0(X,Y1,M).
if0(X,Y,M)$\leftarrow$ Y>=M,X1=X-1, while0(X1,Y,M).
\end{lstlisting}
\end{tabular}

\noindent
The interpreter applied to these clauses can run a goal 
of the form 
\texttt{solve(while0(5,3,10),....)} and terminate. 

 The offline partial evaluator {\sc logen}
 \cite{LeuschelEVCF06} was used to partially evaluate the
 interpreter with respect to a set of clauses and a fixed
 finite set of properties. To use {\sc logen}, each call in the
 interpreter is annotated as \emph{unfold} or \emph{memo}, and
 each argument of memoed calls is annotated as \emph{static},
 \emph{dynamic} or \emph{nonvar} (meaning that everything below
 the top level of the term is dynamic). In the interpreter
 sketched above, the calls to \texttt{solve} and \texttt{delta}
 are memoed, while all other calls, including \texttt{delta_D},
 are unfolded.  The arguments  \texttt{Q}, \texttt{Phi},
 \texttt{Psi} and \texttt{Prog} are \emph{static}, while
 \texttt{A} is  \emph{nonvar}.  The specialised program thus
 consists solely of specialised clauses for
 \texttt{solve} and the concrete constraints linking one
 concrete state with the next. 
 
 \textbf{Example result.} The result of specialization of the clauses 
 above, using the same set of properties in the \texttt{Psi}
 argument as were used in \cite{Gallagher-VPT2019} is as follows.
 
 \begin{tabular}{l|l}
\begin{lstlisting}
solve__2(A,B,C) :- A>0, 
   solve__3(A,B,C).
solve__2(A,B,C) :- A=<0.
solve__3(A,B,C) :-B<C,A>0,
   solve__4(A,B+1,C).
solve__3(A,B,C) :- B>=C,
   solve__5(A-1,B,C).
\end{lstlisting}
~&~
\begin{lstlisting}

solve__4(A,B,C) :- A>0, solve__3(A,B,C).
solve__5(A,B,C) :- A>0,B>=C,
   solve__6(A-1,B,C).
solve__5(A,B,C) :- A=<0.
solve__6(A,B,C) :- B>=C,D=A-1,
   solve__5(D,B,C).
\end{lstlisting}
\end{tabular}

\noindent
This result is identical, apart from predicate names,
to the result obtained in \cite{Gallagher-VPT2019}.
Polyvariance is exemplified by \texttt{solve_2},  \texttt{solve_4} and \texttt{solve_5}, these being three versions of calls to
\texttt{solve} when 
interpreting a call to \texttt{while0} in the input clauses, corresponding to different values for the static arguments (the properties that hold) arising during partial evaluation.
Similarly,
\texttt{solve_3} and \texttt{solve_6} are versions of
the interpretation of \texttt{if0}.
 
 The implementation of \texttt{delta_D} reused code from the specializer described in \cite{Gallagher-VPT2019}, but the abstract interpreter has a simpler structure than the specializer used in that work.  This is due to the fact that the operations handling unfolding, memoing, generalization and polyvariance are handled by {\sc logen} and do not need to be included in the interpreter. Furthermore, it would be simple to replace the code for \texttt{delta_D}  with an implementation of an abstract transfer function for some other domain.

\textbf{Related and future work.}
The transformation of programs by specialising interpreters goes back to the Futamura projections \cite{Futamura}. The projections can be exploited by inserting more sophisticated interpreters between a program and the specializer (e.g.\cite{Gallagher-86,Turchin:93:JFP,GlueckJoergensen:94:SAS,Jones:04:whatnot}). The power of the overall program transformation has been improved by combining specialization with abstraction~\cite{HDL:98:ABPS,Puebla-Hermenegildo-Gallagher:PEPM99,Leuschel:04,DBLP:journals/fuin/FioravantiPPS13}. The
main contrast to previous work on combining specialization with abstract interpretation is that we choose not to integrate abstract interpretation in the specializer, but into the interpreter. Thus a simple partial evaluator (in our case {\sc Logen} can 
achieve the same results as the more elaborate specializers incorporating abstract interpretation.
We argue that the approach of combining the interpretive approach with an abstract interpreter has practical advantages such as modularity and ease of implementation. The same transformation power of a sophisticated specializer can be achieved by interpreter specialization provided the underlying specializer is Jones-optimal and performs static expression reduction~\cite{Glueck:02:bti}.  Often it is easier to modify an interpreter than the specialization tool. Also, it only requires to reason about the correctness of the interpreter provided the underlying specializer is correct. 

The approach presented here needs further research. For instance, interpreters may be parameterised by abstract interpretation domains; the `binding-time improvement' of the interpreter is thereby done only once.
The approach is not limited to offline specialization; other specialization tools such as online specializers and supercompilers may be used. Clearly, the interpretive approach lends itself to generate specializers by the specializer projections~\cite{Glueck:94:JFP}. These will be challenges for further investigations.

\bibliographystyle{eptcs-patched}

\begin{thebibliography}{10}
\providecommand{\bibitemdeclare}[2]{}
\providecommand{\surnamestart}{}
\providecommand{\surnameend}{}
\providecommand{\urlprefix}{Available at }
\providecommand{\url}[1]{\texttt{#1}}
\providecommand{\href}[2]{\texttt{#2}}
\providecommand{\urlalt}[2]{\href{#1}{#2}}
\providecommand{\doi}[1]{doi:\urlalt{http://dx.doi.org/#1}{#1}}
\providecommand{\bibinfo}[2]{#2}

\bibitemdeclare{inproceedings}{Cousot-Cousot}
\bibitem{Cousot-Cousot}
\bibinfo{author}{P.~\surnamestart Cousot\surnameend} \&
  \bibinfo{author}{R.~\surnamestart Cousot\surnameend}
  (\bibinfo{year}{{1977}}): \emph{\bibinfo{title}{Abstract interpretation: a
  unified lattice model for static analysis of programs by construction or
  approximation of fixpoints}}.
\newblock In: {\sl \bibinfo{booktitle}{POPL}}, pp. \bibinfo{pages}{238--252},
  \doi{10.1145/512950.512973}.

\bibitemdeclare{article}{DBLP:journals/tplp/DomenechGG19}
\bibitem{DBLP:journals/tplp/DomenechGG19}
\bibinfo{author}{J.~J. \surnamestart Dom{\'{e}}nech\surnameend},
  \bibinfo{author}{J.~P. \surnamestart Gallagher\surnameend} \&
  \bibinfo{author}{S.~\surnamestart Genaim\surnameend} (\bibinfo{year}{2019}):
  \emph{\bibinfo{title}{Control-Flow Refinement by Partial Evaluation, and its
  Application to Termination and Cost Analysis}}.
\newblock {\sl \bibinfo{journal}{{TPLP}}}
  \bibinfo{volume}{19}(\bibinfo{number}{5-6}), pp. \bibinfo{pages}{990--1005},
  \doi{10.1017/S1471068419000310}.

\bibitemdeclare{article}{DBLP:journals/fuin/FioravantiPPS13}
\bibitem{DBLP:journals/fuin/FioravantiPPS13}
\bibinfo{author}{F.~\surnamestart Fioravanti\surnameend},
  \bibinfo{author}{A.~\surnamestart Pettorossi\surnameend},
  \bibinfo{author}{M.~\surnamestart Proietti\surnameend} \&
  \bibinfo{author}{V.~\surnamestart Senni\surnameend} (\bibinfo{year}{2013}):
  \emph{\bibinfo{title}{Controlling Polyvariance for Specialization-based
  Verification}}.
\newblock {\sl \bibinfo{journal}{Fundam. Inform.}}
  \bibinfo{volume}{124}(\bibinfo{number}{4}), pp. \bibinfo{pages}{483--502},
  \doi{10.3233/FI-2013-845}.

\bibitemdeclare{article}{Futamura}
\bibitem{Futamura}
\bibinfo{author}{Y.~\surnamestart Futamura\surnameend} (\bibinfo{year}{1971}):
  \emph{\bibinfo{title}{Partial Evaluation of Computation Process - An Approach
  to a Compiler-Compiler}}.
\newblock {\sl \bibinfo{journal}{Systems, Computers, Controls}}
  \bibinfo{volume}{2(5)}, pp. \bibinfo{pages}{45--50}.

\bibitemdeclare{inproceedings}{Gallagher-86}
\bibitem{Gallagher-86}
\bibinfo{author}{J.~P. \surnamestart Gallagher\surnameend}
  (\bibinfo{year}{1986}): \emph{\bibinfo{title}{Transforming Logic Programs by
  Specialising Interpreters}}.
\newblock In: {\sl \bibinfo{booktitle}{Proceedings of the 7th European
  Conference on Artificial Intelligence (ECAI-86), Brighton}}, pp.
  \bibinfo{pages}{109--122}.

\bibitemdeclare{inproceedings}{Gallagher-VPT2019}
\bibitem{Gallagher-VPT2019}
\bibinfo{author}{J.~P. \surnamestart Gallagher\surnameend}
  (\bibinfo{year}{2019}): \emph{\bibinfo{title}{Polyvariant program
  specialisation with property-based abstraction}}.
\newblock In \bibinfo{editor}{A.~\surnamestart Lisitsa\surnameend} \&
  \bibinfo{editor}{A.~P. \surnamestart Nemytykh\surnameend}, editors: {\sl
  \bibinfo{booktitle}{VPT-19}}, {\sl \bibinfo{series}{{EPTCS}}}
  \bibinfo{volume}{299}, \doi{10.4204/EPTCS.299.6}.

\bibitemdeclare{article}{Glueck:94:JFP}
\bibitem{Glueck:94:JFP}
\bibinfo{author}{R.~\surnamestart Gl{\"u}ck\surnameend} (\bibinfo{year}{1994}):
  \emph{\bibinfo{title}{On the generation of specializers}}.
\newblock {\sl \bibinfo{journal}{Journal of Functional Programming}}
  \bibinfo{volume}{4}(\bibinfo{number}{4}), pp. \bibinfo{pages}{499--514},
  \doi{10.1017/S0956796800001167}.

\bibitemdeclare{inproceedings}{Glueck:02:bti}
\bibitem{Glueck:02:bti}
\bibinfo{author}{R.~\surnamestart Gl{\"u}ck\surnameend} (\bibinfo{year}{2002}):
  \emph{\bibinfo{title}{{Jones} Optimality, Binding-Time Improvements, and the
  Strength of Program Specializers}}.
\newblock In: {\sl \bibinfo{booktitle}{Proc. Asian Symposium on Partial
  Evaluation and Semantics-Based Program Manipulation}},
  \bibinfo{publisher}{ACM}, pp. \bibinfo{pages}{9--19},
  \doi{10.1145/568173.568175}.

\bibitemdeclare{inproceedings}{GlueckJoergensen:94:SAS}
\bibitem{GlueckJoergensen:94:SAS}
\bibinfo{author}{R.~\surnamestart Gl{\"u}ck\surnameend} \&
  \bibinfo{author}{J.~\surnamestart J{\o}rgensen\surnameend}
  (\bibinfo{year}{1994}): \emph{\bibinfo{title}{Generating transformers for
  deforestation and supercompilation}}.
\newblock In \bibinfo{editor}{B.~\surnamestart Le~Charlier\surnameend}, editor:
  {\sl \bibinfo{booktitle}{Static Analysis. Proceedings}},
  \bibinfo{series}{LNCS 864}, \bibinfo{publisher}{Springer-Verlag}, pp.
  \bibinfo{pages}{432--448}, \doi{10.1007/3-540-58485-4\_57}.

\bibitemdeclare{inproceedings}{HDL:98:ABPS}
\bibitem{HDL:98:ABPS}
\bibinfo{author}{J.~\surnamestart Hatcliff\surnameend},
  \bibinfo{author}{M.~\surnamestart Dwyer\surnameend} \&
  \bibinfo{author}{S.~\surnamestart Laubach\surnameend} (\bibinfo{year}{1998}):
  \emph{\bibinfo{title}{Staging static analyses using abstraction-based program
  specialization}}.
\newblock In \bibinfo{editor}{C.~\surnamestart Palamidessi\surnameend} et~al.,
  editors: {\sl \bibinfo{booktitle}{Principles of Declarative Programming}},
  \bibinfo{series}{LNCS 1490}, \bibinfo{publisher}{Springer}, pp.
  \bibinfo{pages}{134--151}, \doi{10.1007/BFb0056612}.

\bibitemdeclare{book}{Jones-Gomard-Sestoft}
\bibitem{Jones-Gomard-Sestoft}
\bibinfo{author}{N.~D. \surnamestart Jones\surnameend},
  \bibinfo{author}{C.~\surnamestart Gomard\surnameend} \&
  \bibinfo{author}{P.~\surnamestart Sestoft\surnameend} (\bibinfo{year}{1993}):
  \emph{\bibinfo{title}{{P}artial {E}valuation and {A}utomatic {S}oftware
  {G}eneration}}.
\newblock \bibinfo{publisher}{Prentice Hall},
  \doi{10.1016/j.scico.2004.03.010}.

\bibitemdeclare{article}{Jones:04:whatnot}
\bibitem{Jones:04:whatnot}
\bibinfo{author}{N.~D. \surnamestart Jones\surnameend} (\bibinfo{year}{2004}):
  \emph{\bibinfo{title}{Transformation by interpreter specialization}}.
\newblock {\sl \bibinfo{journal}{SCP}}
  \bibinfo{volume}{52}(\bibinfo{number}{1-3}), pp. \bibinfo{pages}{307--339},
  \doi{10.1016/j.scico.2004.03.010}.

\bibitemdeclare{article}{Leuschel:04}
\bibitem{Leuschel:04}
\bibinfo{author}{M.~\surnamestart Leuschel\surnameend} (\bibinfo{year}{2004}):
  \emph{\bibinfo{title}{A framework for the integration of partial evaluation
  and abstract interpretation of logic programs}}.
\newblock {\sl \bibinfo{journal}{ACM TOPLAS}}
  \bibinfo{volume}{26}(\bibinfo{number}{3}), pp. \bibinfo{pages}{413--463},
  \doi{10.1145/982158.982159}.

\bibitemdeclare{inproceedings}{LeuschelEVCF06}
\bibitem{LeuschelEVCF06}
\bibinfo{author}{M.~\surnamestart Leuschel\surnameend},
  \bibinfo{author}{D.~\surnamestart Elphick\surnameend},
  \bibinfo{author}{M.~\surnamestart Varea\surnameend},
  \bibinfo{author}{S.~\surnamestart Craig\surnameend} \&
  \bibinfo{author}{M.~\surnamestart Fontaine\surnameend}
  (\bibinfo{year}{2006}): \emph{\bibinfo{title}{The {Ecce} and {Logen} partial
  evaluators and their web interfaces}}.
\newblock In \bibinfo{editor}{J.~\surnamestart Hatcliff\surnameend} \&
  \bibinfo{editor}{F.~\surnamestart Tip\surnameend}, editors: {\sl
  \bibinfo{booktitle}{PEPM}}, \bibinfo{publisher}{{ACM}}, pp.
  \bibinfo{pages}{88--94}, \doi{10.1145/1111542.1111557}.

\bibitemdeclare{inproceedings}{Puebla-Hermenegildo-Gallagher:PEPM99}
\bibitem{Puebla-Hermenegildo-Gallagher:PEPM99}
\bibinfo{author}{G.~\surnamestart Puebla\surnameend},
  \bibinfo{author}{M.~\surnamestart Hermenegildo\surnameend} \&
  \bibinfo{author}{J.~P. \surnamestart Gallagher\surnameend}
  (\bibinfo{year}{1999}): \emph{\bibinfo{title}{An integration of partial
  evaluation in a generic abstract interpretation framework}}.
\newblock In \bibinfo{editor}{O.~\surnamestart Danvy\surnameend}, editor: {\sl
  \bibinfo{booktitle}{PEPM'99}}, \bibinfo{address}{San Antonio, Texas}, pp.
  \bibinfo{pages}{75--84}.

\bibitemdeclare{article}{Turchin:93:JFP}
\bibitem{Turchin:93:JFP}
\bibinfo{author}{V.~F. \surnamestart Turchin\surnameend}
  (\bibinfo{year}{1993}): \emph{\bibinfo{title}{Program transformation with
  metasystem transitions}}.
\newblock {\sl \bibinfo{journal}{Journal of Functional Programming}}
  \bibinfo{volume}{3}(\bibinfo{number}{3}), pp. \bibinfo{pages}{283--313},
  \doi{10.1017/S0956796800000757}.

\end{thebibliography}

\end{document}